\documentclass[conference]{IEEEtran}
\makeatletter
\def\ps@headings{%
\def\@oddhead{\mbox{}\scriptsize\rightmark \hfil \thepage}%
\def\@evenhead{\scriptsize\thepage \hfil \leftmark\mbox{}}%
\def\@oddfoot{}%
\def\@evenfoot{}}
\makeatother
\IEEEoverridecommandlockouts
\usepackage{cite}
\usepackage{amsmath,amssymb,amsfonts}
\usepackage{algorithmic}
\usepackage{multicol}
\usepackage{graphicx}
\usepackage{float}
\usepackage{listings}
\usepackage{textcomp}
\usepackage{xcolor}
\def\BibTeX{{\rm B\kern-.05em{\sc i\kern-.025em b}\kern-.08em
    T\kern-.1667em\lower.7ex\hbox{E}\kern-.125emX}}
    
\usepackage{booktabs} % For formal tables
\usepackage{multirow} % For merged cell in tables
\usepackage{diagbox} % For backslash column in tables
\usepackage{array}
\newcolumntype{x}[1]{>{\centering\arraybackslash}p{#1}}
\newcolumntype{y}[1]{>{\arraybackslash}p{#1}}
 
\usepackage[left=0.65in, right=0.65in, top=0.825in, bottom=1in]{geometry}

\begin{document}

\title{PhishZip: A New Compression-based Algorithm for ﻿Detecting Phishing Websites}
%\title{PhishZip: Detecting Phishing Websites by Compressing Them}

\author{
    \IEEEauthorblockN{Rizka Purwanto\IEEEauthorrefmark{1}\IEEEauthorrefmark{3}, Arindam Pal\IEEEauthorrefmark{2}\IEEEauthorrefmark{3}, Alan Blair\IEEEauthorrefmark{1}, Sanjay Jha\IEEEauthorrefmark{1}}
    \IEEEauthorblockA{
    \IEEEauthorrefmark{1}\textit{School of Computer Science and Engineering, University of New South Wales, Australia}
    }
    \IEEEauthorblockA{
    \IEEEauthorrefmark{2}\textit{Data61, CSIRO, Sydney, New South Wales, Australia}
    }
    \IEEEauthorblockA{
    \IEEEauthorrefmark{3}\textit{Cyber Security Cooperative Research Centre, Australia}
    }
}

% \author{\IEEEauthorblockN{Rizka Purwanto}
% \IEEEauthorblockA{\textit{School of Computer Science and Engineering} \\
% \textit{University of New South Wales}\\
% Sydney, Australia \\
% r.purwanto@unsw.edu.au
% }
% \and
% \IEEEauthorblockN{Arindam Pal}
% \IEEEauthorblockA{\textit{Data61} \\
% \textit{CSIRO}\\
% Sydney, Australia \\
% arindam.pal@data61.csiro.au}
% \and
% \IEEEauthorblockN{Alan Blair}
% \IEEEauthorblockA{\textit{School of Computer Science and Engineering} \\
% \textit{University of New South Wales}\\
% Sydney, Australia \\
% a.blair@unsw.edu.au}
% \and

% \IEEEauthorblockN{Sanjay Jha}
% \centerline{
% \IEEEauthorblockA{\textit{School of Computer Science and Engineering} \\
% \textit{University of New South Wales}\\
% Sydney, Australia \\
% sanjay.jha@unsw.edu.au}}
% }

\maketitle

\begin{abstract}
%%This document is a model and instructions for \LaTeX.
%%This and the IEEEtran.cls file define the components of your paper [title, text, heads, etc.]. *CRITICAL: Do Not Use Symbols, Special Characters, Footnotes, 
%%or Math in Paper Title or Abstract.
Phishing has grown significantly in the past few years and is predicted to further increase in the future. The dynamics of phishing introduce challenges in implementing a robust phishing detection system and selecting features which can represent phishing despite the change of attack. In this study, we propose PhishZip which is a novel phishing detection approach using a compression algorithm to perform website classification and demonstrate a systematic way to construct the word dictionaries for the compression models using word occurrence likelihood analysis. PhishZip outperforms the use of best-performing HTML-based features in past studies, with a true positive rate of 80.04\%. We also propose the use of compression ratio as a novel machine learning feature which significantly improves machine learning based phishing detection over previous studies. Using compression ratios as additional features, the true positive rate significantly improves by 30.3\% (from 51.47\% to 81.77\%), while the accuracy increases by 11.84\% (from 71.20\% to 83.04\%).
\end{abstract}


\section{Introduction}
\label{sec:introduction}

The number of phishing attacks has grown significantly in the past few years. The Anti-Phishing Working Group (APWG) recorded a significant increase of unique phishing attacks from 2014 to 2016 \cite{apwg2016} causing considerably high financial loss ranging between \$60 million and \$3 billion per year in the United States \cite{Hong2012}. The number of attacks is likely to increase in the future with the availability of phishing toolkits and algorithms which ease the process of phishing \cite{sood2016taxonomy, holz2008measuring}. The dynamics in phishing behaviours bring challenges in implementing a robust and accurate phishing detection for the long term \cite{Ma:2009:ISU:1553374.1553462}.

%While there are plenty of studies in spam and phishing detection, the challenge remains to effectively deal with the dynamic behaviour of phishing \cite{Ma:2009:ISU:1553374.1553462}. Due to this dynamic, the detection system should have the ability to evolve and select features which are still relevant despite the change of attack. One possible solution is by extracting higher level features in the textual content to capture certain patterns instead of extracting features from the surface level text \cite{Gutierrez2018SAFEPC}. However, text-based features are sensitive to changes made by phishers. For example, an attacker could intentionally use different terms or make incorrect spelling to bypass phishing detection systems. Therefore, it would be ideal to have a mechanism to frequently and flexibly update the keywords representing phishing websites.

To mitigate the negative impacts of phishing, security software providers, financial institutions, and academic researchers have studied various approaches to build an automated phishing website detection system. These methods include the use of blacklists and detecting phishing websites by investigating the website content, URL, and web-related features \cite{Xiang2011, xiang2009hybrid, Zhang2007cantina, zhang2011textual, Whittaker2010}.

While most past studies in phishing detection focused on the use of machine learning algorithms, this study introduces the use of compression algorithms to perform phishing website classification. The compression-based classification process aims to predict whether a certain website is phishing or benign based on the textual information. To perform this task, we built two compression models which are optimised for phishing and benign websites respectively. Building these models requires information regarding the word distribution in phishing and benign websites, which is required to optimise the compression for each class. Using these models, we can perform classification by comparing the compression ratios of both models. We expect that a phishing optimised model should compress phishing websites better than a non-phishing optimised compression model, resulting in a higher compression ratio, and vice versa. The main tool that we use in this work is the \texttt{zlib} library, which is based on the DEFLATE compression algorithm \cite{RFC1951}. DEFLATE is a lossless data compression algorithm used for compressing websites in HTTP \cite{RFC2616}.

To summarise, this paper makes the following contributions:
\begin{itemize}
    \item We introduce a systematic process of selecting meaningful words which are associated with phishing and non-phishing websites by analysing the likelihood of word occurrences and calculate the optimal likelihood threshold. These set of words are used as the predefined compression dictionary for our compression models. %and are also useful as features to perform classification using machine learning models.
    \item We develop a tool called PhishZip which performs phishing website detection using the DEFLATE compression algorithm. To the best of our knowledge, this study is the first to use compression algorithms to perform phishing website classification. Unlike machine learning based models, performing classification by leveraging compression algorithms does not require training the models nor performing HTML parsing \cite{marton2005compression}. Thus, classification with compression algorithms is faster and simpler. %To obtain a good classification performance, we need prior information regarding the word distribution in the phishing websites.
    \item We propose the use of compression ratio as a novel machine learning feature which is robust and easy to extract. Compression ratio measures the distance or cross-entropy between the predicted website and phishing/non-phishing website content distribution. The high compression ratio is associated with low cross-entropy, which indicates that the content distribution is similar to the common word distribution in phishing/non-phishing websites \cite{marton2005compression}. 
    %\item We analyzed the likelihood of word occurrences in each class and the optimal likelihood threshold to select words to be included as the DEFLATE preset dictionary for our compression models. This set of words are also useful as features to perform phishing classification using machine learning models, which are shown by its classification performance, e.g. accuracy and F1 score. Thus, we can perform word analysis to formally derive text-based features for machine learning based phishing detection models.
    % \item We performed a statistical analysis which shows a positive correlation between the compression ratio and the website content size for each compression models. Based on this result, we attempt to adjust our detection model by applying a certain weight to penalise compression ratio values of websites with relatively large content size. This adjustment empirically leads to improvement of the F1-score of our compression based model.
    %\item We evaluate the compression based and machine learning based detection models performances. The evaluation metrics include accuracy, false positive rate (FPR), and F1-score.
\end{itemize}

%%%[Paper structure]
The paper is organised as follows. In section \ref{sec:related}, we provide an overview of past studies in phishing detection systems and compression based classification. Section \ref{sec:background} provides some background regarding phishing and the concept of compression-based classification. Section \ref{sec:system} gives the system overview of PhishZip. We provide the word occurrence likelihood analysis for constructing the word dictionary of the compression models in Section \ref{sec:word}. The experimental setups, including the evaluation methodology and diversity of our web page corpus, are presented in Section \ref{sec:experiments}. We provide the performance evaluation results in Section \ref{sec:results} and provide more analysis regarding these results in Section \ref{sec:discussion}. Finally, we wrap up with conclusions in Section \ref{sec:conclusion}.

\section{Related Works}
\label{sec:related}
Several past studies in phishing website detection systems have been conducted, mainly focusing on the use of machine learning algorithms to classify phishing websites based on specific features. Whittaker et al. introduced a machine learning classifier for automatically maintaining Google's phishing blacklist using features extracted from the URL, page hosting information, and the page content \cite{Whittaker2010, Gutierrez2018SAFEPC}. Meanwhile, Xiang et al. proposed CANTINA+ as a comprehensive framework for detecting phishing websites using URL-based, HTML-based, and web-based features \cite{Xiang2011}. The study by Xiang et al. extends CANTINA, which performs phishing detection based on TF-IDF information retrieval algorithm and seven other content-based heuristics, including age of domain, logo image and domain name inconsistency, and suspicious links in the HTML \cite{Zhang2007cantina}. Xiang and Hong also studied the use of identity discovery as a hybrid phishing detection approach \cite{xiang2009hybrid}. Meanwhile, Zhang et al. \cite{zhang2011textual} proposed a content-based phishing website detection by analysing the website textual and visual content and assessing its similarity. In a recent study, Quinkert et al. \cite{quinkert2019s} focused on observing the use of homograph domains to identify scamming and phishing.

In contrast, our work leverages compression algorithm to perform classification which has not been implemented for phishing detection in past studies. A number of studies have previously discussed the use of data compression algorithm to perform text classification in various areas. Marton et al. in \cite{marton2005compression} evaluated the performance of various compression-based classification methods to classify text based on topic/genre and authorship attribution. Meanwhile, Ziegelmayer and Schrader in \cite{ziegelmayer2012sentiment} discussed the use of Prediction by Partial Matching (PPM) to perform sentiment polarity classification. Compression algorithms have also been used to perform classification of amino acid sequences of DNA as discussed in \cite{chiba2001classification}.

\section{Background}
\label{sec:background}
In this section, we provide background on phishing, compression-based classification, and a brief description of the DEFLATE compression algorithm.

\subsection{Phishing}
Phishing is a social engineering attack which aims at stealing user account credentials by impersonating legitimate organisations or institutional websites. Several motives underlie these attacks, such as financial gain or stealing identities for hiding illegal actions \cite{6497928}. To perform phishing, attackers typically broadcast emails to the victims with urgent calls to a particular action, e.g. account subscription status or password reset due to a data breach, with a URL redirecting to these phishing websites. As this type of attack exploits users' vulnerabilities, finding an effective strategy to avoid phishing and reduce its impact is fairly challenging. Despite having an information system which is technically secure against password theft, attackers are still able to obtain user data when unaware end-users type their credentials into a phishing website form.

%The detection system extracts features from phishing data, then uses various machine learning algorithms to identify phishing. One of the features commonly used is textual content, which is extracted using specific Natural Language Processing (NLP) and information retrieval techniques, e.g. word frequency analysis and TF-IDF (term frequency-inverse document frequency) extraction \cite{Zhang2007cantina, Gutierrez2018SAFEPC}. With an effective extraction process, textual information is intuitively useful to provide keywords representing the main website content, which could indicate whether a website is associated with phishing attempts.

Three types of anti-phishing techniques exist to mitigate the negative impacts of phishing \cite{Abu-Nimeh2007}, which are detection, prevention, and revision. The detection technique is arguably the most effective as it minimises human errors, the main vulnerability exploited by phishing attackers \cite{Dou2017}. Developing a phishing detection system requires some knowledge regarding how attackers conduct the attacks. In their literature study, Dou et al. mentioned five stages of phishing attacks as reconnaissance, weaponisation, distribution, exploitation, and exfiltration \cite{Dou2017}. Based on this life-cycle, phishing detection systems are implemented to avoid phishing during the exploitation step, when victims receive the phishing emails. The detection systems detect phishing websites or phishing emails on the client side through a Web browser or specific anti-phishing software, or on the server side. Whenever a phishing website or email is detected, the detection system will either block user access to the suspected website, notify users through a warning that the visited website or received email is potentially malicious \cite{varshney2016survey, fette2007learning}.

\subsection{Compression-Based Classification}

Text categorisation tasks assign each document to one of the preset categories. Various machine learning algorithms have been used in several studies to perform text categorisation tasks, such as Naive Bayes, SVM, and deep learning methods. Machine learning based text classification typically consists of four steps: (1) word and sentence segmentation on the training files, (2) feature selection based on the word counts, (3) building a model based on a machine learning algorithm, (4) performing feature extraction on the testing files and evaluating the model on this test data \cite{teahan2003using}.

Compared to the standard machine learning approach for classification, compression-based methods are relatively easy to apply and do not require further input document pre-processing or feature extraction \cite{teahan2003using}. Data compression algorithms build a model of the processed documents based on extensive statistics regarding these documents \cite{ziegelmayer2012sentiment}. Text classification is performed by compressing the documents using compression models optimised for each class of document. After obtaining the compression results, each document is assigned to the preset category which results in the highest compression rate, indicating the best compression. From the perspective of information theory, the compression rate is related to the cross-entropy between the distribution of the class text corpus and the text distribution of the predicted document. By choosing the class model which performs the best for the predicted document, the classification is essentially performed by choosing the category whose text corpus minimises this cross-entropy \cite{marton2005compression}. In this study, we leverage a compression algorithm as a novel approach to perform phishing website classification. Further details regarding the DEFLATE algorithm that we use are provided in the following subsection.

% For this study, we use DEFLATE lossless compression algorithm which has already been implemented in Python standard library as \texttt{zlib} \cite{zlib}.

% In this study, we use the DEFLATE lossless compression algorithm which is popular for Web compression. DEFLATE is one of the compression schemes used in the HTTP standard for minimising latency by reducing the byte size transferred across the network \cite{RFC2616}. As one of the Web compression algorithms defined in the HTTP protocol, DEFLATE compression and decompression are standard features in Web browsers; thus, no additional package or add-on installation is required \cite{microsoftCompress}. This opens the opportunity for implementing phishing website detection feature on Web browsers which makes use of the existing Web compression scheme, which will be discussed further in the next section regarding our proposed system.

\subsection{DEFLATE Algorithm}
In this study, we use the DEFLATE lossless compression algorithm which is one of the compression schemes used in the HTTP standard for minimising latency by reducing the byte size transferred across the network \cite{RFC2616}. In Web browsers, two of the three encodings defined in the HTTP specification ("\texttt{Content-Encoding: gzip}" and "\texttt{Content-Encoding: deflate}") are based on the DEFLATE algorithm \cite{microsoftCompress}.

%DEFLATE is one of the compression schemes used in the HTTP standard for minimising latency by reducing the byte size transferred across the network \cite{RFC2616}. DEFLATE algorithm is a combination of the LZ77 algorithm and Huffman coding as specified in RFC 1951 \cite{RFC1951}. 

The DEFLATE algorithm mainly consists of the LZ77 algorithm and Huffman encoding. LZ77 is a dictionary-based Lempel-Ziv compression algorithm which aims to eliminate duplicate bytes by replacing recurring bytes in the data with a back-reference which points to the first occurrence of the bytes. The next step is using Huffman coding for replacing symbols with new weighted symbols depending on the frequency of occurrence. Thus, symbols which frequently show up are replaced with shorter symbol representations, and vice versa \cite{deflateAlgorithm}.

We have used the \texttt{zlib} data compression library which provides an abstraction of the DEFLATE compression algorithm \cite{zlib}. The \texttt{zlib} module in Python includes functions which implement data compression and decompression using the DEFLATE algorithm \cite{zlibPython} and provides the functionality to set a predefined zlib compression dictionary \cite{RFC1950}. This word dictionary can help to improve the compression result and should contain a list of bytes or strings which are expected to occur frequently in the data \cite{zlibPython}. In this study, we built two word-dictionaries, which consist of common words in phishing and non-phishing websites respectively. With these two dictionaries, we performed compression on the same document \emph{twice} using each dictionary separately. Ideally, compressing phishing website content using the phishing dictionary should produce a more compressed output than using the non-phishing dictionary and vice versa.

\section{System Overview}
\label{sec:system}
This section contains a description regarding the design of PhishZip detection system. There are two main modules in PhishZip, which are the dictionary builder module and compression model.

\subsection{Dictionary Builder}
This module aims at creating word dictionaries to build compression models optimised for phishing websites and non-phishing websites respectively. To build these dictionaries, we calculate the likelihood of words occurring in each website class (i.e. phishing and non-phishing) and selecting a minimum likelihood threshold value. The words in the dictionaries are those whose likelihood of occurring in a certain website class are greater than a certain threshold.

The word occurrence likelihood in a specific website class is proportional to the word frequency in this class; thus, words which show up more often will be assigned higher likelihood values. Meanwhile, this likelihood value should be comparable for both website classes. Thus, for an arbitrary word, if its likelihood value of occurrence in phishing websites is higher than in non-phishing websites, we should expect the word to appear more often in phishing websites, which indicates higher association to phishing websites than non-phishing websites.

To estimate the likelihood of a word $w_k$ occurring in a certain text category $v_j \in \{phishing, non-phishing\}$ (either \textit{'phishing'} or \textit{'non-phishing'}), we adopt the \textit{m}-estimate with uniform priors and set $m = |V|$ (the size of the word vocabulary $V$). More formally,
\begin{equation}\label{eq:likelihood}
    P(w_k|v_j) = \frac{n_k+1}{n+|V|}.
\end{equation}

Here, $n$ is the total number of word positions in all training examples whose target value is $v_j$, $n_k$ is the number of times the word $w_k$ is found in all the $n$ word positions, and $|V|$ is the total number of distinct words and other tokens in all examples \cite{Mitchell:1997:ML:541177}.

\subsection{Compression-Based Classification Model}
% The Python zlib library implements the DEFLATE compression algorithm and handles the compression process in our work. In this study, the compression objects are set with the default compression level (6), which provides the default compromise between the compression speed and performance. The \texttt{wbit} parameter, which represents the base-two logarithm of the window size, is set to -15, giving the maximum setting window size ($2^{15}$ bits) while producing a raw output without header or trailing checksum.

To perform the classification task, we build two compression models optimised for phishing websites and non-phishing websites respectively. The phishing optimised compression model uses the phishing dictionary, while the non-phishing optimised compression model uses the non-phishing dictionary as the preset compression dictionary. These dictionaries are constructed from the dictionary builder process as discussed in the previous subsection.

To perform website classification, we compress the raw website HTML content using the phishing optimised model and non-phishing optimised model separately. There are two separate compression outputs from this process. We compare these outputs by calculating the compression ratio, which is the ratio of the original input size to the compression output size.

\begin{equation}\label{eq:compression_ratio}
    Compression~Ratio_i = \frac{size(HTML\_content)}{size(C_i(HTML\_content))}.
\end{equation}

\begin{figure}
    \centering
    \includegraphics[width=0.5\textwidth]{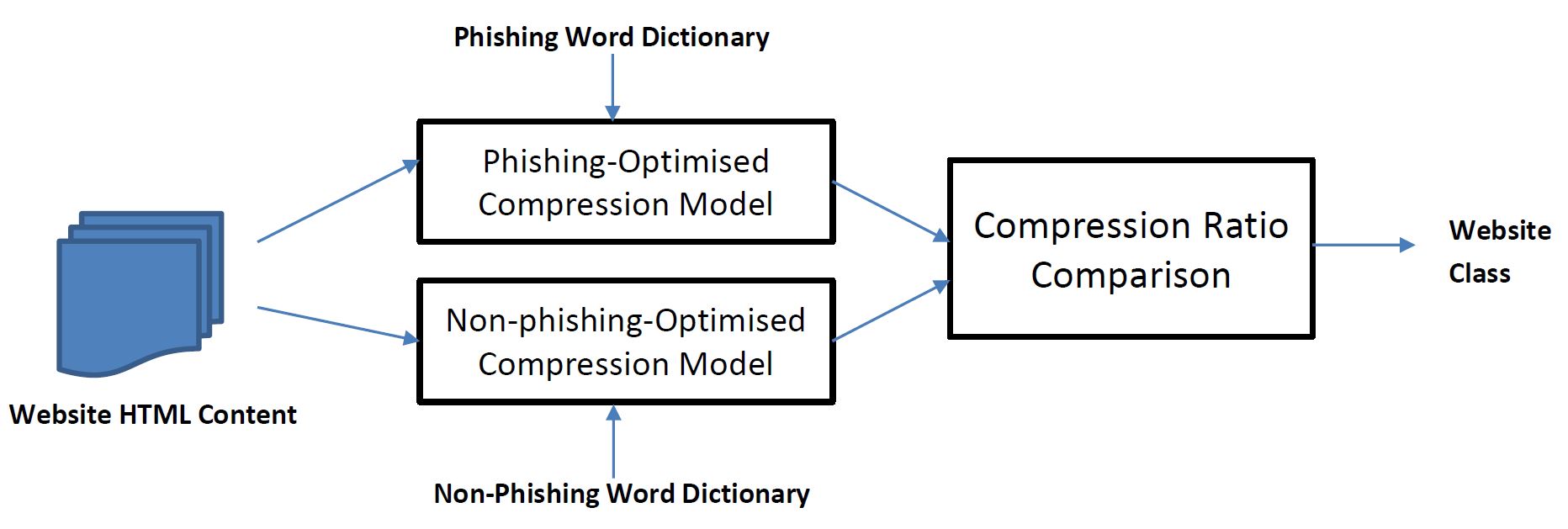}
    \caption{PhishZip Compression-Based Classification Model}
    \label{fig:phishzhip_compression_model}
\end{figure}

Here, \texttt{$C_i$} indicates the compression model and \texttt{HTML\_content} is the raw website content. The decision on whether the website is classified as phishing or non-phishing is based on which compression model produces the higher compression ratio, i.e. if the phishing optimised compression model produces a better compression ratio than the non-phishing optimised model, then the website is classified as phishing, and vice versa. An illustration of the compression-based classification model is shown in Figure~\ref{fig:phishzhip_compression_model}.
 
% \begin{lstlisting}[language=Python, frame=single, basicstyle=\footnotesize]
% if compress_ratio_phish > compress_ratio_nonphish:
%     website_cat = 'phishing'
% else:
%     website_cat = 'nonphishing'
% \end{lstlisting}

\section{Word Occurrence Likelihood Analysis}
\label{sec:word}
The compression-based classification mainly relies on word distribution differences in each class. Thus, we performed analysis to observe the word distribution in the website textual contents and calculate the likelihood of word occurrences in phishing and non-phishing websites. This information is useful to build the dictionary of common words in phishing and non-phishing websites for the compression models. We also performed an analysis to select the optimal likelihood threshold for constructing the common word dictionary which leads to the best classification performance. Before performing these analyses, we performed some data preprocessing to obtain the website textual contents. These processes will be described further in the following subsections.

To perform this analysis, we use a set of website HTML contents which consist of 5,000 phishing and 5,000 non-phishing websites. These website contents were manually collected by fecthing the content of phishing website URLs from PhishTank [22] and safe website URLs listed by Quantcast [6] using \texttt{curl} \cite{curl} to securely inspect the website contents without running any potentially malicious code on the browser.

% This Web scraping process was performed using a command line based tool for fetching website content named \texttt{curl} \cite{curl} to securely inspect the website contents without running any potentially malicious code on the browser.

As we focus on the textual information in the website, we performed data preprocessing to extract text from the website HTML content and removed common English stop words (e.g. "the", "a", "an") as specified in Natural Language Toolkit (NLTK) library \cite{nltk}. After performing these data preprocessing steps, we obtained a collection of phishing and non-phishing website textual contents which were used to perform word occurrence likelihood analysis.

\begin{figure}
    \centering
    \includegraphics[width=0.45\textwidth]{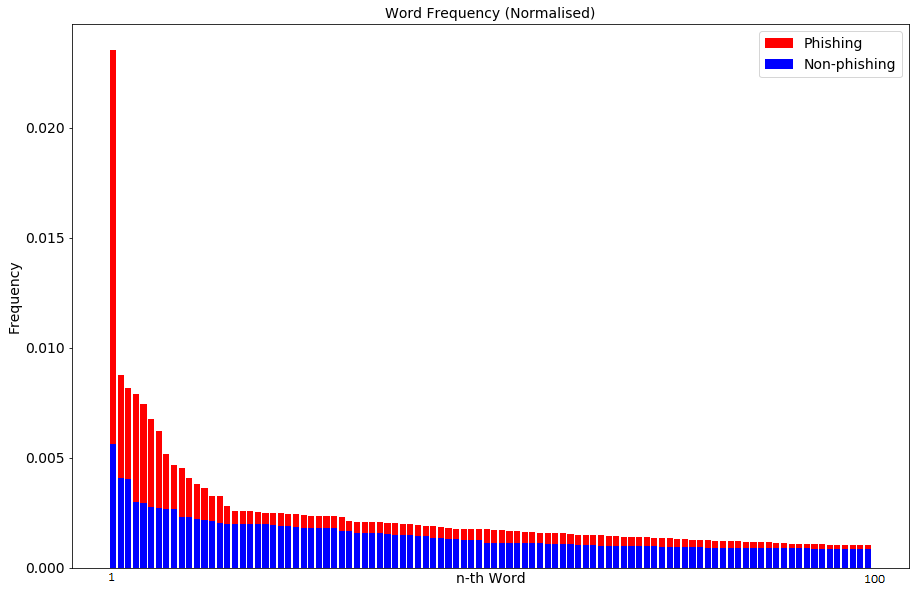}
    \caption{Word Frequency Histogram}
    \label{fig:word_hist_norm_B}
\end{figure}

To observe the word distribution in both classes and how distinctive they are, we plot the frequency of words in both phishing and non-phishing text corpora. We normalised the word frequency by dividing the value by the length of phishing or non-phishing text corpus to show the frequency of word occurrence relative to the total number of words in the corpus. The histograms of word distribution of phishing and non-phishing websites are shown in Figure~\ref{fig:word_hist_norm_B}. In this histogram, we included the 100 most frequent words to show the word distribution differences more clearly. The histogram shows that the word distribution of phishing websites is typically much steeper than non-phishing websites. This intuitively means that phishing websites typically share common words and use less variety of words in the website content. Meanwhile, the use of words in non-phishing websites are more general as shown by the flatter distribution shape.

After calculating word occurrence likelihood calculation using Equation~\ref{eq:likelihood}, likelihood threshold analysis is performed to obtain the likelihood threshold which leads to the best classification accuracy. This optimisation is done by varying the likelihood threshold and creating a hypothetical predefined dictionary using the words whose likelihood of occurrence in phishing and non-phishing websites is greater than this threshold. Two compression models are built using the phishing and non-phishing predefined word dictionaries. Afterwards, classification is performed by comparing which model results in a smaller file output size. The accuracy is calculated for each likelihood threshold.

As the word list size is enormous (536,684 unique words in the phishing text corpus and 3,668,395 unique words in the non-phishing text corpus), for practical reasons, we only stored 3,000 words with the highest likelihood in each text corpus and its likelihood values. To obtain a general idea of how many words are included in the dictionary as we vary the likelihood thresholds, we plot the distribution of the likelihood values. This plot could also help in analysing the accuracy value dynamics and fluctuations when we vary the likelihood thresholds in a specific range. 

\begin{figure}
    \centering
    \includegraphics[width=0.35\textwidth]{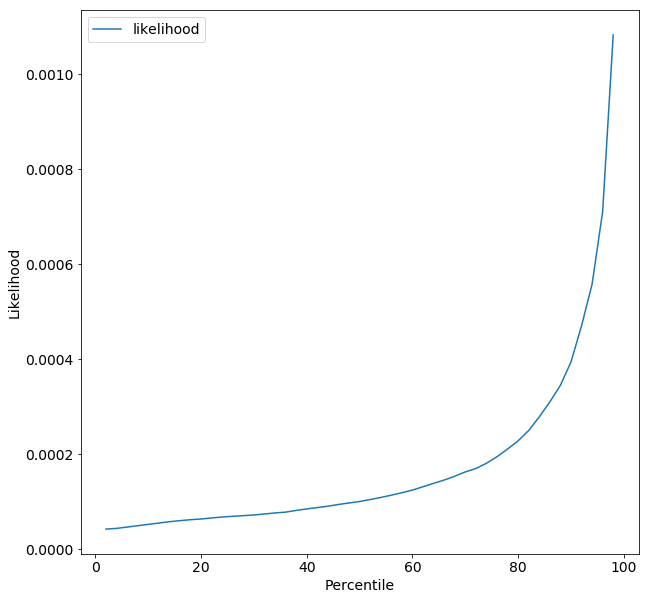}
    \caption{Distribution of Word Occurrence Likelihood}
    \label{fig:word_likelihood_dist_B}
\end{figure}

Figure~\ref{fig:word_likelihood_dist_B} shows the word occurrence likelihood values of the $10^{th}$, $20^{th}$, up to the $100^{th}$ percentile. The graph shows that for both tasks, the likelihood value increases exponentially. This distribution shows that the majority of word occurrence likelihood (around 80\% of words in the lower-rank) ranges between ~0.00001 to ~0.0002. Meanwhile, there is a steep increase in likelihood values in the top 20\% of words. The optimal likelihood threshold would eliminate as many words as possible, but still include words which are meaningful in providing information regarding a category.

% The optimal likelihood threshold would eliminate as many words as possible which are not quite meaningful in providing information regarding a category, but at the same time still including words which have a meaningful representation of the categories.

\begin{figure}
    \centering
    \includegraphics[width=0.45\textwidth]{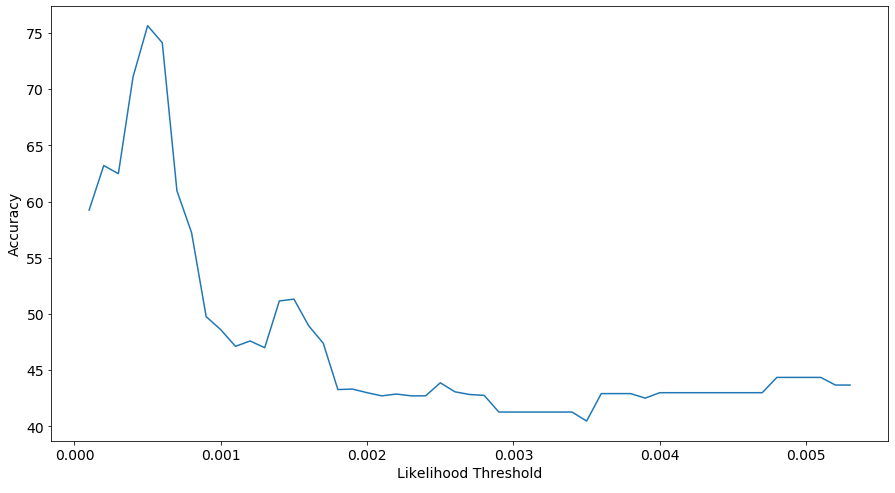}
    \caption{Accuracy vs Likelihood Threshold}
    \label{fig:accuracy_vs_likelihood_thr_B}
\end{figure}

Meanwhile, Figure~\ref{fig:accuracy_vs_likelihood_thr_B} shows the accuracy trend as the likelihood threshold is varied. This graph shows a relatively regular pattern where the accuracy increases as the likelihood threshold increases and gradually decreases after reaching the optimal likelihood threshold value. The maximum accuracy (75.64\%) is reached when the likelihood threshold is set to 0.0005. From the distribution of word occurrence likelihood (Figure~\ref{fig:word_likelihood_dist_B}), 0.0005 is around the $94^{th}$ percentile in the likelihood data. Thus, by setting the likelihood threshold to 0.0005, we are selecting the top 6\% of words from our list which are associated with phishing and non-phishing classes. Setting this likelihood threshold results in a phishing word dictionary which consists of 178 words and a non-phishing word dictionary that contains 246 words. Sample of words with high likelihood values in both dictionaries are provided in Table~\ref{tab:word_cloud}. We also attempted to select the words using other methods, i.e., based on the occurrence likelihood difference of phishing and non-phishing websites. However, we were unable to achieve good performances with the dictionaries constructed using this word selection method, as the method includes words with high likelihood differences and low occurrence likelihoods into the dictionaries; thus, adding words which do not frequently appear in non-phishing or phishing websites.

\begin{table}
\caption{Sample of Words in the Dictionaries}
\label{tab:word_cloud}
 \begin{center}
  \begin{tabular}{x{3cm}x{3cm}}
   \toprule
   Phishing &
   Non-phishing\\
   \midrule
        email & us\\
        account & news\\
        sign & get\\
        password & view\\
        please & free\\
        server & best\\
        help & shop\\
        deleting & day\\
   \bottomrule
  \end{tabular}
 \end{center}
\end{table}

\section{Experimental Setup}
\label{sec:experiments}
In this section, we briefly provide the evaluation methodology that we use in this study, including the performance metrics to assess the models' performances, followed by further details regarding the Web page corpus size and distribution.

\subsection{Evaluation Methodology}
We conducted two experiments in this study. In the first experiment, we examined the performance of PhishZip in detecting phishing websites and compared this to a machine learning model using HTML-based features proposed by Xiang et al. in CANTINA+ \cite{Xiang2011}. Meanwhile in the second experiment, we evaluated the performance of compression ratios as machine learning features and investigated how this could improve the detection performance of CANTINA+. In these experiments, we focus on the following metrics to evaluate the performances:
\begin{itemize}
    \item True positive rate (TPR): the ratio between the number of correctly classified phishing websites and the total number of phishing websites,
    \item False positive rate (FPR): the ratio between the number of misclassified non-phishing websites and the total number of non-phishing websites.
    \item F1-score: the harmonic mean of recall (true positive rate) and precision.
\end{itemize}

\subsection{Web Page Corpus}
The Web page corpus used to perform these performance evaluations consist of 2,045 phishing and 2,000 non-phishing raw HTML contents. To avoid classification model overfitting, this Web page collection is an entirely different website dataset used for performing word likelihood analysis and building the compression dictionaries.

To contrast, the phishing website dataset for word likelihood analysis comprises of 5,000 phishing websites chosen randomly which were reported by users to PhishTank \cite{phishTank} from 2016 to 2018 (665 websites in 2016, 1921 websites in 2017, and 2414 websites in 2018). On the other hand, the phishing website dataset used for the performance evaluations consists of phishing websites reported to PhishTank between January to May 2019. While we were able to collect 8,492 verified phishing websites listed by PhishTank, 6,447 of these websites have unknown target brand (labeled as 'Others' in PhishTank) and only 2,045 of these are provided with information of the legitimate target brand they attempt to resemble. To strictly observe the dataset diversity, we excluded phishing websites whose target brand name are undefined. The total number of target brands in this dataset is 92 with PayPal, Facebook, and Microsoft as the top brand targets. Further details regarding the phishing dataset diversity are provided in Table~\ref{tab:phishing_target}.

\begin{table}
\caption{Top 10 Phishing Target Brands}
\label{tab:phishing_target}
 \begin{center}
  \begin{tabular}{x{3cm}x{3cm}}
   \toprule
   Brand Name &
   Number of Websites\\
   \midrule
        Paypal & 846 \\
        Facebook & 262 \\
        Microsoft & 125 \\
        ABSA Bank & 117 \\
        RuneScape & 107 \\
        ING Direct & 79 \\
        eBay, Inc. & 56 \\
        Delta Air Lines & 40 \\
        Allegro & 30 \\
        Blockchain & 25 \\
   \bottomrule
  \end{tabular}
 \end{center}
\end{table}

Meanwhile, the non-phishing website dataset for this performance analysis consists of 2,000 non-phishing Web pages selected at random from the list of safe websites by Quantcast \cite{quantCast}. Similar to our approach when building the phishing dataset, we also make sure that the non-phishing websites in this dataset are not one of the websites included in the non-phishing dataset for the word likelihood analysis.

\section{Results}
\label{sec:results}
In this section, we provide the results of two experiments conducted in this study: the first experiment examines the performance of PhishZip as a compression based classification method for detecting phishing websites, while the second experiment investigates the performance of website compression ratio as a viable machine learning feature for phishing detection systems.

\subsection{PhishZip Performance Evaluation}
To assess the performance of PhishZip, we compress each website in the phishing and non-phishing evaluation dataset with compression models optimised for phishing and non-phishing websites respectively, then compare the compression ratio of each model. Phishing websites are expected to have higher compression ratio using phishing optimised compression model than non-phishing compression model, and vice versa.

By performing classification using compression algorithms alone, we are able to detect phishing websites with a true positive rate of 80.04\%, false positive rate of 18.25\%, and accuracy of 80.89\%.

% Further details regarding the number of true positives and true negatives from the phishing classification are provided in Table~\ref{tab:phishzip_confusion_matrix}.

% \begin{table}
% \caption{PhishZip Confusion Matrix}
% \label{tab:phishzip_confusion_matrix}
%  \begin{center}
%     \begin{tabular}{l|l|c|c|}
%     \multicolumn{2}{c}{} & \multicolumn{2}{c}{Actual} \\
%     \cline{3-4}
%     \multicolumn{2}{c|}{} & Phishing & Non-phishing \\
%     \cline{2-4}
%     \multirow{2}{*}{Prediction} & Phishing & 1637 & 365 \\
%     \cline{2-4}
%     & Non-phishing & 408 & 1635 \\
%     \cline{2-4}
%     \end{tabular}
%  \end{center}
% \end{table}

% \begin{table*}
% \caption{PhishZip Performance Comparison to Other HTML-Based Heuristics}
% \label{tab:phishzip_performance_comparison}
%  \begin{center}
%   \begin{tabular}{x{3cm}x{2cm}x{2cm}x{3cm}x{3cm}}
%   \toprule
%   Performance Metrics &
%   PhishZip & 
%   Bad forms &
%   Bad action fields &
%   Non-matching URLs \\
%   \midrule
%     TPR & 80.04\% & 8.89\% & 19.95\% & 47.92\% \\
%     FPR & 18.25\% & 19.45\% & 38.25\% & 8.00\% \\
%     Accuracy & 80.89\% & 44.32\% & 40.62\% & 69.71\% \\
%     F1-score & 80.89\% & 13.91\% & 25.36\% & 61.54\% \\
%   \bottomrule
%   \end{tabular}
%  \end{center}
% \end{table*}

\begin{table}
\caption{PhishZip Performance Comparison to CANTINA+}
\label{tab:phishzip_performance_comparison}
 \begin{center}
  \begin{tabular}{x{2cm}x{2cm}x{3cm}}
  \toprule
  Performance Metrics &
  PhishZip & 
  CANTINA+ (HTML-based features)\\
  \midrule
    TPR & 80.04\% & 51.47\% \\
    FPR & 18.25\% & 8.92\% \\
    Accuracy & 80.89\% & 71.20\% \\
    F1-score & 80.89\% & 64.21\% \\
  \bottomrule
  \end{tabular}
 \end{center}
\end{table}

As a comparison, we also perform phishing website classification with features proposed by Xiang et al. in CANTINA+ \cite{Xiang2011}. We choose the features proposed in this study as they provide comprehensive results regarding the performance of the system and each individual feature. In their study of CANTINA+, Xiang et al. investigated the performance of various URL-based features, HTML-based features, and Web-based features. The top-performing features are page in top search results, bad forms, bad action fields, and non-matching URLs. As our work focuses on website content based features, we compare the performance of PhishZip with top-performing HTML-based features only, which are bad forms, bad action fields, and non-matching URLs.

A summary of the performance evaluation results of PhishZip and the selected features in CANTINA+ is shown in Table~\ref{tab:phishzip_performance_comparison}. We found that the machine learning model achieved an accuracy of 71.20\%, low false positive rate around 8.92\%, and relatively low true positive rate of roughly 51.47\%. These results show that while the use of these HTML-based features still produce in a relatively descent accuracy and low false positive rate, they did not perform well in detecting phishing websites as reflected by the true positive rate. Meanwhile, as shown in Table~\ref{tab:phishzip_performance_comparison}, PhishZip outperforms the use of HTML-based features in CANTINA+ with a true positive rate of 80.04\%, false positive rate of 18.25\%, and accuracy of 80.89\%. Further results of each HTML-based feature are discussed as follows.

\subsubsection{Bad forms}
One of the best performing HTML-based features in CANTINA+ \cite{Xiang2011} is the bad form indicator, which is a binary feature set to 1 if any malicious form exist in the Web page HTML content. A malicious form exists if a Web page has all of the following:
\begin{itemize}
    \item an HTML form,
    \item an \textless{}input\textgreater{} tag in the form,
    \item sensitive keywords (e.g. "password", or "credit card number") or image only (with no text) in the scope of the form,
    \item a non-https URL in the action field of the form or in the Web page URL (if the action form is empty).
\end{itemize}

We applied the approach above to identify bad forms in the 2,045 phishing and 2,000 non-phishing Web page corpora. Using this heuristic, we categorise the website as phishing if any bad forms exist, and vice versa. This approach gives us a true positive rate of 8.90\% and false positive rate of 19.45\%.

%The exact numbers of the true positives and the true negatives from this classification are provided in the confusion matrix in Table~\ref{tab:bad_forms_confusion_matrix}.

% \begin{table}
% \caption{Confusion Matrix (Bad Forms Feature)}
% \label{tab:bad_forms_confusion_matrix}
%  \begin{center}
%     \begin{tabular}{l|l|c|c|}
%     \multicolumn{2}{c}{} & \multicolumn{2}{c}{Actual} \\
%     \cline{3-4}
%     \multicolumn{2}{c|}{} & Phishing & Non-phishing \\
%     \cline{2-4}
%     \multirow{2}{*}{Prediction} & Phishing & 182 & 389 \\
%     \cline{2-4}
%     & Non-phishing & 1863 & 1611 \\
%     \cline{2-4}
%     \end{tabular}
%  \end{center}
% \end{table}

While the true negative rate is relatively descent (80.55\%), plenty of the phishing websites (91.10\% from the phishing Web page corpora) were misclassified as benign. This is due to the strict condition that the bad forms should have a non-https scheme URL in the action field or the Web page URL. We found that 1,209 out of 2,045 phishing websites (59.12\%) in our phishing Web page corpora use HTTPS in its URL. Further, Anti-Phishing Working Group (APWG) \cite{apwg2016} reported that there is a significant increase in the use of secure connection in phishing websites to increase its credibility.

\subsubsection{Bad action fields}
Another best performing features in CANTINA+ \cite{Xiang2011} is the bad action field indicator, which is a binary feature that detects if an action field is empty or a simple file name, or if it points to a different domain than the Web page domain. Assessing the binary feature performance on our Web page corpora, this approach was able to achieve a true positive rate of 19.95\% and false positive rate of 38.25\%. Further, we found that 232 phishing websites in our corpora have empty or simple filename in the action field and 182 phishing websites have a domain in the action field different to the Web page domain.

% The confusion matrix in Table~\ref{tab:bad_action_fields_confusion_matrix} shows the number of correctly classified phishing and non-phishing websites.

% \begin{table}
% \caption{Confusion Matrix (Bad Action Fields Feature)}
% \label{tab:bad_action_fields_confusion_matrix}
%  \begin{center}
%     \begin{tabular}{l|l|c|c|}
%     \multicolumn{2}{c}{} & \multicolumn{2}{c}{Actual} \\
%     \cline{3-4}
%     \multicolumn{2}{c|}{} & Phishing & Non-phishing \\
%     \cline{2-4}
%     \multirow{2}{*}{Prediction} & Phishing & 408 & 765 \\
%     \cline{2-4}
%     & Non-phishing & 1637 & 1235 \\
%     \cline{2-4}
%     \end{tabular}
%  \end{center}
% \end{table}

\subsubsection{Non-matching URLs}
The other best performing HTML-based feature in CANTINA+ \cite{Xiang2011} is the non-matching URLs indicator, which is set to 1 if the percentage of highly similar URLs or the percentage of empty or ill-formed URL in the website is greater than a certain threshold. To obtain this threshold, we observed the distribution of these percentages and chose the threshold which gives the best performance in terms of the total number of correctly classified samples (true positive + true negative). Using this setting, we are able to detect phishing websites using this heuristic with a true positive rate of 47.92\% and a relatively low false positive rate of 8.0\%.

\subsection{Compression Ratios as Machine Learning Features}
Compression ratio is a novel machine learning feature. To the best of our knowledge, this has not been applied to detect phishing websites in past studies. Compression ratio is also relatively robust against the dynamic behaviour of phishing, as we could easily update the predefined word dictionaries and optimise the compression models correspondingly. Calculating compression ratio is relatively easy, as website browsers perform website compression frequently to improve user experience when opening websites \cite{microsoftCompress}. As defined in HTTP specification, compression is one of the options in the standard for minimising latency during data transfer over the internet \cite{RFC2616} and is a built-in feature in current Web browsers \cite{microsoftCompress}. This could be incorporated through protocol option support, where no re-compression may be needed. These characteristics make compression ratio as a viable option as a machine learning feature for performing phishing detection.

\subsubsection{Dataset Size}
We allocated roughly 80\% of our Web page corpus (1,672 phishing websites and 1,630 non-phishing websites) for training the machine learning model and around 20\% (373 phishing websites and 370 non-phishing websites) to evaluate the performance of the machine learning model. We apply temporal-split between training and testing dataset, following past studies that recommend this approach over a randomised cross-validation evaluation which introduces performance overestimate due to the risk of training future data and testing on the past \cite{ho2019detecting, pendlebury2019tesseract}. Our phishing training dataset consists of phishing websites reported to PhishTank in January-April 2019 targeting 86 unique brands, while the testing dataset are phishing websites reported to PhishTank in May 2019. Meanwhile, we collected the non-phishing Web page corpus in a single time point rather than collecting at each time point following the methodology of CANTINA+ \cite{Xiang2011}, which is based on the study by Fetterly et al. discovering that Web page content is relatively stable over time \cite{Fetterly:2003:ECN:951953.952397}.

\subsubsection{Model Training}
We used several machine learning algorithms to perform classification, including logistic regression, support vector machine, k-nearest neighbours, decision tree, random forest, neural network, and Naive Bayes. We used the \texttt{scikit-learn} library in Python which provides the implementation of these machine learning algorithms. Model parameter tuning is performed by using grid search to find the model parameters which produce the best performance. We used 3-fold cross-validation method to validate the model performance during the model training process which is the default cross-validation method of \texttt{GridSearchCV} in the \texttt{scikit-learn} library.

\subsubsection{Performance Evaluation}

\begin{table}
\caption{Machine Learning Model Performance using Compression Ratios as Features}
\label{tab:result_comp_ratio_ml}
 \begin{center}
  \begin{tabular}{x{2.5cm}x{1cm}x{1cm}x{1cm}x{1.2cm}}
   \toprule
   \multirow{2}{*}{Classifier} &
   \multicolumn{4}{c}{Performance} \\
   \cline{2-5}
   & \textit{TPR}
   & \textit{FPR}
   & \textit{Accuracy}
   & \textit{F1-score} \\
   \midrule
    Logistic regression & 75.34\% & 17.03\% & 79.14\% & 78.38\%\\
    SVM & 78.82\% & 16.76\% & 81.02\% & 80.66\%\\
    K-nearest neighbours & 77.48\% & 12.43\% & 82.50\% & 81.64\%\\
    Decision tree & 75.34\% & 11.89\% & 81.70\% & 80.52\%\\
    \textbf{Random forest} & \textbf{78.28}\% & \textbf{12.97}\% & \textbf{82.64}\% & \textbf{81.91}\%\\
    Neural network & 79.89\% & 30.27\% & 74.83\% & 76.11\%\\
    Naive Bayes & 42.09\% & 8.38\% & 66.76\% & 55.97\%\\
   \bottomrule
  \end{tabular}
 \end{center}
\end{table}

Using solely compression ratios as features of our machine learning model, the best performance is achieved by the model trained using random forest algorithm with an F1-score of 81.91\%. The random forest algorithm works well in this classification problem as it combines different individual models which lead to less bias and less variance. The neural network is able to achieve the best true positive rate of 79.89\%; however, this model suffers from a high false positive rate. The performances of the machine learning models are provided in Table~\ref{tab:result_comp_ratio_ml}.

% The model trained using k-nearest neighbours algorithm also achieves similar performances as the random forest model, with true positive rate of 77-79\%, false positive rate of 12-13\%, accuracy of 82-83\%, and F1-score of 81-82\%.

% We compared the performance of compression ratios as machine learning with the best performing HTML-based features proposed in CANTINA+ \cite{Xiang2011}, namely bad forms, bad action fields, and non-matching URLs. Using the same dataset and model training approach, we found the machine learning models achieved similar results with an accuracy of roughly 71.20\%, low false positive rate around 8.92\%, and relatively low true positive rate of roughly 51.47\%. These results show that while the use of these HTML-based features still produce in a relatively descent accuracy and low false positive rate, they did not perform well in detecting phishing websites as reflected by the true positive rate.

We also assessed the performance of the use of compression ratios and HTML-based features in CANTINA+ to perform phishing website detection. Using both these features, the best performing model achieved a true positive rate of 81.77\%, false positive rate of 15.68\%, and accuracy of 83.04\%. This shows that the combination of compression ratios and HTML-based features could enhance the performance of the phishing website detection with improvements in the true positive rate of 58.87\% over the model using HTML-based features in CANTINA+. While there is a compromise in an increase of false positive rate, the use of this combination of feature has also helped in improving the accuracy and F1-score as well. A comparison of the best performances using various combination of features is provided in Table~\ref{tab:result_ml_performance_comparison}.

% The best performance was achieved by the support vector machine based model with a true positive rate of 81.77\%, false positive rate of 15.68\%, and accuracy of 83.04\%.

\begin{table}
\caption{Machine Learning Model Performance Comparison}
\label{tab:result_ml_performance_comparison}
 \begin{center}
  \begin{tabular}{x{1.3cm}x{1.35cm}x{1.55cm}x{2.55cm}}
   \toprule
   \multirow{2}{1.3cm}{Performance metrics} &
   \multicolumn{3}{c}{Features} \\
   \cline{2-4}
   & \textit{Compression ratios}
   & \textit{HTML-based features}
   & \textit{Compression ratios \& HTML-based features} \\
   \midrule
   TPR & 78.28\% & 51.47\% & 81.77\% \\
   FPR & 12.97\% & 8.92\% & 15.68\% \\
   Accuracy & 82.64\% & 71.20\% & 83.04\% \\
   F1-score & 81.91\% & 64.21\% & 82.88\% \\
   \bottomrule
  \end{tabular}
 \end{center}
\end{table}

We also attempted to assess the performance using an imbalanced testing dataset with a class ratio of non-phishing to phishing websites of 100:1, as suggested in past studies in phishing detection systems \cite{Dou2017, Whittaker2010}. During this evaluation, we down-sampled the phishing dataset by selecting a few phishing samples in random and iterate the testing process \textit{n} number of times using different random phishing samples. In this experiment, we chose $n = 100$ iterations and calculated the average true positive rate, false positive rate, accuracy, and F1-score. A summary of the machine learning models performances in this experiment is provided in Table~\ref{tab:result_ml_performance_comparison_imba}.

Using HTML-based features in CANTINA+, the best model was able to achieve an accuracy of 90.65\%. However, accuracy would not be the best metric to measure the model performance in an imbalanced dataset. As shown in Table~\ref{tab:result_ml_performance_comparison_imba}, the model only achieved a true positive rate of 51\%, which means around half of the phishing websites in the dataset are not detected. As the class ratio of non-phishing to phishing is 100:1, misclassifying one phishing website would still lead to 99\% accuracy. Thus, in this scenario, we measure the model performance based on the true positive rate. Combining the HTML-based features with compression ratios, the best model was able to achieve a significant increase of true positive rate, from 51\% to 82.50\%.

\begin{table}
\caption{Machine Learning Model Performance Comparison (Imbalanced Dataset)}
\label{tab:result_ml_performance_comparison_imba}
 \begin{center}
  \begin{tabular}{x{1.3cm}x{1.35cm}x{1.55cm}x{2.55cm}}
  \toprule
  \multirow{2}{1.3cm}{\centering Performance metrics} &
  \multicolumn{3}{c}{Features} \\
  \cline{2-4}
  & \textit{Compression ratios}
  & \textit{HTML-based features}
  & \textit{Compression ratios \& HTML-based features} \\
  \midrule
  TPR & 77.25\% & 51.00\% & 82.50\% \\
  FPR & 12.97\% & 8.92\% & 15.68\% \\
  Accuracy & 86.92\% & 90.65\% & 84.30\% \\
  F1-score & 11.17\% & 10.34\% & 10.08\% \\
  %Precision & 6.02\% & 5.75\% & 5.37\% \\
  \bottomrule
  \end{tabular}
 \end{center}
\end{table}

\section{Discussion}
\label{sec:discussion}
In this section, we discuss the model evaluation results in Section \ref{sec:results}, including some limitations of the approach and possible future works.

\subsection{Performance Evaluation}
In this study, we propose the use of compression algorithm to perform phishing website classification. The compression ratio measures the cross entropy between the distribution of the website content we are trying to classify and the word distribution of phishing or non-phishing websites. As shown in Table~\ref{tab:phishzip_performance_comparison}, PhishZip outperforms the performance of top-performing HTML-based features proposed in CANTINA+ \cite{Xiang2011}, achieving a true positive rate of 80.04\%. The use of compression ratios as an additional feature to add with CANTINA+ features is able to significantly improve the machine learning model performance as shown in Table~\ref{tab:result_ml_performance_comparison}, achieving a true positive rate of 81.77\% and accuracy of 83.04\%. As shown in Table~\ref{tab:result_ml_performance_comparison}, using solely compression ratios, we are also able to detect phishing websites with a comparably good true positive rate of 78.28\% and an accuracy of 82.64\%. These results show that compression ratio is a viable option as a machine learning feature for detecting phishing websites and as an additional feature to improve the performance of existing phishing website detection systems.

\subsection{Limitations and Future Works}
There are some limitations in the current PhishZip implementation. First, it may not perform well when classifying websites which are purely made up of images without any text to analyse. Therefore, the system can possibly be bypassed when attackers use images only on the phishing website pages to imitate the design of the targeted website. On the other hand, legitimate websites rarely contain solely images and no text \cite{Xiang2011}. This characteristic would be useful to distinguish legitimate websites from phishing websites. One possible solution is by training classification models to detect potential phishing Web pages with no sufficient text and mostly filled up with images.

PhishZip also suffers when dealing with websites with content obfuscation, e.g. using an external file to load the website content, as well as cross site scripting (XSS) attacks and the use of \texttt{iframe}. However, we believe that existing methods, e.g. cookie protection and DOM or HTML sanitizer, would complement PhishZip and provide solutions to mitigate these vulnerabilities.

% \begin{figure*}
%     \centering
%     \includegraphics[width=0.6\textwidth]{images/website_obfuscation.png}
%     \caption{Website Content Obfuscation}
%     \label{fig:web_content_obfuscation}
% \end{figure*}

There are also possibilities during which attackers intentionally avoid PhishZip detection by selecting words which are not included in the compression dictionaries or using words with low likelihood of showing up in phishing websites. One possible approach to avoid this scenario is by keeping the likelihoods and dictionaries confidential. This is similar to the case of machine learning models, where the model architecture, weights, or parameters should be kept secure to avoid adversarial attacks.

For future works, there is also an opportunity to improve the performance of PhishZip by making use of website HTML structure information. As discussed by Cui et al. \cite{cui2017tracking}, phishing websites have similar HTML DOMs and are often variations of other phishing websites. This information might be useful to optimise the compression models better or select the best compression algorithms which will be well-suited to compress websites with slight variation of the HTML DOM trees.

\section{Conclusion}
\label{sec:conclusion}
In this study, we propose PhishZip as a novel approach to perform phishing website classification using a dictionary-based compression algorithm. This method leverages word dictionaries constructed by analysing the word occurrence likelihood, which is also demonstrated in this work. PhishZip outperforms the use of best-performing HTML-based features proposed in past studies with a true positive rate of 80.04\%. We also introduced the use of compression ratio as a novel machine learning feature which has shown to significantly improve past studies in machine learning based phishing detection systems. Using compression ratios as additional features, the true positive rate has significantly improved by around 30.3\%, from 51.47\% to 81.77\%, while the accuracy increased roughly by 11.84\%, from 71.20\% to 83.04\%.

%%From this study, we performed a phishing classification task using DEFLATE lossless compression algorithm. We experimented with two kinds of dictionaries, word-based and HTML token based dictionary, which consist of words or HTML tokens with relatively high likelihood to occur in phishing and benign websites. Using this method, we are able to classify phishing websites in general with an accuracy value of 75.64\% and false positive rate of 13.44\%. This method could also be implemented to perform phishing sign-in pages classification with an accuracy value of 68.4\% and false positive rate of 26.4\%, which exceeds the baseline performance by ZeroR classifier (accuraccy=50\%) by 18.4\%.


\section*{Acknowledgment}
\label{sec:acknowledgment}

The work has been supported by the Cyber Security Research Centre Limited whose activities are partially funded by the Australian Government’s Cooperative Research Centres Programme.

Rizka Widyarini Purwanto was supported by a UNSW University International Postgraduate Award (UIPA) scholarship. Any opinions, findings, and conclusions or recommendations expressed in this paper are those of the authors and do not necessarily reflect the views of the scholarship provider.

\bibliographystyle{IEEEtran}
\bibliography{references}

\end{document}